\documentclass[twocolumn,showpacs,preprintnumbers,amsmath,amssymb,nofootinbib,floatfix]{revtex4}

\usepackage[dvips]{graphicx}
\usepackage{dcolumn}
\usepackage{bm}
\usepackage{braket}

\def\Journal#1#2#3#4{{#1} {\bf #2}, #3 (#4)}
 
\def\AandA{Astron. Astrophys.}

\def\CPC{Chin. Phys. C}
\def\EPJC{Eur. Phys. J. C}
\def\IJMPA{Int. J. Mod. Phys. A}

\def\JCAP{J. Cosmol. Astropart. Phys.}
\def\JHEP{J. High Energy Phys.}

\def\JPG{J. Phys. G} 
\def\MPLA{Mod. Phys. Lett. A}

\def\NJP{New. J. Phys.}
\def\NPB{Nucl. Phys. B}

\def\PLB{{Phys. Lett.} B}

\def\PRL{Phys. Rev. Lett.}
\def\PRD{Phys. Rev. D}

\begin{document}


 \title{Scotogenic dark matter and single-zero textures of the neutrino mass matrix}

\author{Teruyuki Kitabayashi}
\email{teruyuki@tokai-u.jp}

\affiliation{%
\sl Department of Physics, Tokai University, 4-1-1 Kitakaname, Hiratsuka, Kanagawa 259-1292, Japan
}


\begin{abstract}
The scotogenic model can simultaneously account for the presence of dark matter and the origin of neutrino masses. We assume that the flavor neutrino mass matrix has one zero element and Yukawa matrix elements are real in the scotogenic model. It turns out that only one pattern of the flavor neutrino mass matrix in the one-zero-texture scheme within the scotogenic model is viable with the observed neutrino oscillation data, the relic abundance of the dark matter, and the upper limit of the branching ratio of the $\mu \rightarrow e \gamma$ process.
\end{abstract}

\maketitle


\section{Introduction\label{sec:introduction}}
Understanding the nature of both dark matter and neutrinos is one of the big problems in cosmology and particle physics. The scotogenic model, or radiative seesaw model, can simultaneously account for the presence of dark matter and the origin of neutrino masses \cite{Ma2006PRD}. In this model, neutrino masses are generated by one-loop interactions mediated by a dark matter candidate. One-loop interactions related to dark matter and neutrino mass have been extensively studied in the literature \cite{one_loop_Ma1998PRL,one_loop_Kubo2006PLB,one_loop_Hambye2007PRD,one_loop_Farzan2009PRD,one_loop_Farzan2010MPLA,one_loop_Farzan2011IJMPA,one_loop_Kanemura2011PRD,one_loop_Schmidt2012PRD,one_loop_Faezan2012PRD,one_loop_Aoki2012PRD,one_loop_Hehn2012PLB,one_loop_Bhupal2012PRD,one_loop_Bhupal2013PRD,one_loop_Law2013JHEP,one_loop_Kanemura2013PLB,one_loop_Hirsch2013JHEP,one_loop_Restrepo2013JHEP,one_loop_Lindner2014PRD,one_loop_Okada2014PRD89,one_loop_Okada2014PRD90,one_loop_Brdar2014PLB,Toma2014JHEP,Vicente2015JHEP,one_loop_Borah2015PRD,one_loop_Wang2015PRD,one_loop_Fraser2016PRD,one_loop_Adhikari2016PLB,one_loop_Ma2016PLB,one_loop_Arhrib2016JCAP,one_loop_Okada2016PRD,one_loop_Ahriche2016PLB,one_loop_Lu2016JCAP,one_loop_Cai2016JHEP,Suematsu2009PRD,Suematsu2010PRD,Ibarra2016PRD,Lindner2016PRD,Das2017PRD,Singirala2017CPC,Kitabayashi2017IJMPA,Rojas2018arXiv,Baumholzer2018JHEP,Ahriche2018PRD,Hugle2018PRD}.  

One of the key ingredients in the scotogenic model is the Yukawa matrix $Y$. In order to obtain any phenomenological prediction in the scotogenic model, the elements of the Yukawa matrix should be determined. This matrix is closely connected with the neutrino sector. There are some ways to determine the Yukawa matrix elements:

\begin{description}
\item[(a)] Assume an appropriate form of the Yukawa matrix $Y$ (see, e.g., Ref.\cite{Ibarra2016PRD}).
\item[(b)] Use an appropriate parameterization of neutrino mixing to derive the most general form of the Yukawa matrix compatible with the neutrino oscillation data (see, e.g., Ref.\cite{Ibarra2016PRD}).
\item[(c)] Assume an appropriate form of the neutrino mixing matrix $U$ (see, e.g., Ref.\cite{Suematsu2009PRD,Suematsu2010PRD, Singirala2017CPC}).
\item[(d)] Assume an appropriate form of the flavor neutrino mass matrix $M_\nu$ (see, e.g., Ref.\cite{Kitabayashi2017IJMPA}).
\end{description}

In this study, we employ methods (c) and (d). More concretely:
\begin{description}
\item[(c')] We assume that the neutrino mixing matrix $U$ is described as a modified tribimaximal mixing pattern.  The exact tribimaximal pattern is approximately consistent with the observed solar and atmospheric neutrino mixings. However, the exact tribimaximal pattern predicts a vanishing reactor neutrino mixing angle. We know that the reactor neutrino mixing angle is small but moderately large. We employ the modified tribimaximal mixing pattern from Refs.\cite{Singirala2017CPC,Sruthilaya2015NJP}.

\item[(d')] We assume that the flavor neutrino mass matrix $M_\nu$ has one zero element. There have been various discussions on ways to ensure the appearance of the observed neutrino mixings and masses based on flavor neutrino mass matrices with zeros \cite{Ludl2014JHEP}. This type of flavor neutrino mass matrix consists of what are called texture zeros. The origin of such texture zeros was discussed in Refs.\cite{Berger2001PRD,Low2004PRD,Low2005PRD,Grimus2004EPJC,Xing2009PLB,Dev2011PLB,Araki2012JHEP,Felipe2014NPB,Grimus2005JPG}. In paricular, the phenomenology of one-zero and two-zero textures was studied in Refs.\cite{Xing2004PRD,Lashin2012PRD,Deepthi2012EPJC,Gautam2015PRD} and \cite{Cebola2015PRD,Frampton2002PLB,Xing2002PLB530,Xing2002PLB539,Kageyama2002PLB,Dev2007PRD,Ludle2012NPB,Kumar2011PRD,Fritzsch2011JHEP,Meloni2013NPB,Meloni2014PRD,Dev2014PRD,Dev2015EPJC,Kitabayashi2016PRD}, respectively. Also, the experimental potential of probing the texture-zero models has been discussed. For example, the possibility of probing different texture-zero neutrino flavor mass matrices at the long-baseline neutrino experiment DUNE was shown in Ref.\cite{Bora2017PRD}. 

Scenarios with one or two texture zeros for the Yukawa matrix $Y$ in the scotogenic model were studied in, e.g., Ref.\cite{Ibarra2016PRD}. We discuss the possible scenarios with one texture zero for the flavor neutrino mass matrix $M_\nu$ instead of $Y$.  
\end{description}

In this paper, all elements of the Yukawa matrix are taken as real for simplicity. We show that only one pattern of the flavor neutrino mass matrix in the one-zero-texture scheme is viable within the scotogenic model.

This paper is organized as follows. In Sec.\ref{sec:scotogenic}, we present a brief review of the scotogenic model. In Sec.\ref{sec:tribi}, according to method ${\rm (c')}$, we assume that the neutrino mixing matrix is described as a modified tribimaximal mixing pattern \cite{Singirala2017CPC,Sruthilaya2015NJP}. In Sec.\ref{sec:onezero}, according to method ${\rm (d')}$, we employ the one-zero-texture scheme. We show that only one pattern of the flavor neutrino mass matrix in the one-zero-texture scheme within the scotogenic model is consistent with the observed neutrino oscillation data, the relic abundance of dark matter, and the upper limit of the branching ratio of the $\mu \rightarrow e \gamma$ process from analytical and numerical calculations. Section \ref{sec:summary} is devoted to a summary.

\section{Scotogenic model\label{sec:scotogenic}}
The scotogenic model \cite{Ma2006PRD} is an extension of the standard model. This model has three extra Majorana $SU(2)_L$ singlets $N_k$ $(k=1,2,3)$ and one new scalar $SU(2)_L$ doublet $(\eta^+,\eta^0)$. These new particles are odd under exact $Z_2$ symmetry. Under $SU(2)_L \times U(1)_Y \times Z_2$, the main particle content in the scotogenic model is given by  $(\alpha=e,\mu,\tau)$ 
\begin{eqnarray}
&& L_\alpha=(\nu_\alpha, \ell_\alpha)_L \ : \ (2,-1/2,+), \quad \ell_\alpha^C \ : \ (1,1,+),\nonumber \\
&& \Phi=(\phi^+, \phi^0) \ : \ (2,1/2,+), \nonumber \\
&& N_k \ : \  (1,0,-), \quad \eta=(\eta^+,\eta^0) \ : \ (2,1/2,-),  
\end{eqnarray}
where $(\nu_\alpha, \ell_\alpha)$ is the left-handed lepton doublet and $(\phi^+, \phi^0)$ is the Higgs doublet in the standard model. 

The Lagrangian of the scotogenic model contains new terms for the new singlets, 
\begin{eqnarray}
\mathcal{L} \supset Y_{\alpha k} (\bar{\nu}_{\alpha L} \eta^0 - \bar{\ell}_{\alpha L} \eta^+) N_k + \frac{1}{2}M_k \bar{N}_k N^C_k + H.c.,
\label{Eq:L_yukawa}
\end{eqnarray}
and the scalar potential of the model contains the  quartic scalar interaction
\begin{eqnarray}
V \supset \frac{1}{2}\lambda (\Phi^\dagger \eta)^2 + H.c.
\end{eqnarray}
Owing to the $Z_2$ symmetry, neutrinos remain massless at tree level but acquire masses via one-loop interactions. The neutrino flavor mass matrix reads \cite{Ma2006PRD}
\begin{eqnarray}
M_\nu=\left( 
\begin{array}{*{20}{c}}
M_{ee} & M_{e\mu} & M_{e\tau} \\
- & M_{\mu\mu} &M_{\mu\tau}\\
- & - & M_{\tau\tau} \\
\end{array}
\right),
\label{Eq:Mnu}
\end{eqnarray}
where the symbol ``$-$" denotes a symmetric partner. The flavor neutrino masses are obtained as
\begin{eqnarray}
M_{\alpha\beta} = \sum_{k=1}^3 Y_{\alpha k}Y_{\beta k} \Lambda_k,
\label{Eq:M_alpha_beta}
\end{eqnarray}
where
\begin{eqnarray}
\Lambda_k &=&  \frac{\lambda v^2}{16\pi^2}\frac{M_k}{m^2_0-M^2_k}\left(1-\frac{M^2_k}{m^2_0-M^2_k}\ln\frac{m_0^2}{M^2_k} \right),
\label{Eq:Lambda_k} \\
m_0^2 &=& \frac{1}{2}(m_R^2+m_I^2),
\end{eqnarray}
and $v$, $m_R$, and $m_I$ denote the vacuum expectation value of the Higgs field, and the masses of $\sqrt{2} {\rm Re}[\eta^0]$ and $\sqrt{2} {\rm Im}[\eta^0]$, respectively. 

In this model, flavor-violating processes such as $\mu \rightarrow e \gamma$ are induced at the one-loop level. The branching ratio of $\mu \rightarrow e \gamma$  is given by \cite{one_loop_Kubo2006PLB}

\begin{eqnarray}
{\rm Br}(\mu \rightarrow e \gamma)=\frac{3\alpha_{\rm em}}{64\pi(G_Fm_0^2)^2}\left| \sum_{k=1}^3 Y_{\mu k}Y_{e k}^* F \left( \frac{M_k}{m_0}\right) \right|^2,
\end{eqnarray}
where $\alpha_{\rm em}$ denotes the fine-structure constant, $G_F$ denotes the Fermi coupling constant and  $F(x)$ is defined by
\begin{eqnarray}
F(x)=\frac{1-6x^2+3x^4+2x^6-6x^4 \ln x^2}{6(1-x^2)^4}.
\end{eqnarray}

The scotogenic model predicts the existence of a dark matter particle. The lightest $Z_2$-odd particle is stable in the particle spectrum. This lightest $Z_2$-odd particle becomes a dark matter candidate. We know that if we take the coannihilation effect into account \cite{Suematsu2009PRD,Suematsu2010PRD}, the predicted cold dark matter abundance as well as the branching ratio of the lepton-flavor-violating $\mu \rightarrow e \gamma$ process can be simultaneously consistent with observations within the simplest (original) scotogenic model \cite{Ma2006PRD}. We assume that the lightest Majorana singlet fermion, $N_1$, becomes the dark matter particle and $N_1$ is considered to be almost degenerate with the next-to-lightest Majorana singlet fermion $N_2$. In this case, $M_1 \lesssim M_2 < M_3$, we could take account of coannihilation effects \cite{Griest1991PRD}. 

The (co)annihilation cross section times the relative velocity of annihilation particles $v_{\rm rel}$ is given by \cite{Suematsu2009PRD}
\begin{eqnarray}
\sigma_{ij} |v_{\rm rel}|= a_{ij} + b_{ij} v_{\rm rel}^2,
\end{eqnarray}
with
\begin{eqnarray}
a_{ij}&=& \frac{1}{8\pi}\frac{M_1^2}{(M_1^2+m_0^2)^2} \sum_{\alpha\beta}(Y_{\alpha i} Y_{\beta j} - Y_{\alpha j} Y_{\beta i})^2, \nonumber \\
b_{ij}&=&\frac{m_0^4-3m_0^2M_1^2-M_1^4}{3(M_1^2+m_0^2)^2}a_{ij} \nonumber \\
&& +  \frac{1}{12\pi}\frac{M_1^2(M_1^4+m_0^4)}{(M_1^2+m_0^2)^4}  \sum_{\alpha\beta}Y_{\alpha i} Y_{\alpha j} Y_{\beta i} Y_{\beta j},
\label{Eq:a_b}
\end{eqnarray}
where $\sigma_{ij}$ $(i,j=1,2)$ is the annihilation cross section for $N_i N_j \rightarrow \bar{f}f$, $\Delta M = (M_2-M_1)/M_1$ depicts the mass-splitting ratio of the degenerate singlet fermions, $x = M_1/T$ denotes the ratio of the mass of the lightest singlet fermion to the temperature $T$, and $g_1$ and $g_2$ are the number of degrees of freedom of $N_1$ and $N_2$, respectively. The effective cross section $\sigma_{\rm eff}$ is obtained as
\begin{eqnarray}
\sigma_{\rm eff} &=& \frac{g_1^2}{g_{\rm eff}^2}\sigma_{11} + \frac{2g_1g_2}{g_{\rm eff}^2}\sigma_{12} (1+\Delta M)^{3/2}e^{-\Delta M \cdot x}\nonumber \\
&&+ \frac{g_2^2}{g_{\rm eff}^2}\sigma_{22} (1+\Delta M)^3 e^{-2\Delta M \cdot x},\nonumber \\
g_{\rm eff}&=&g_1+g_2 (1+\Delta M)^{3/2}e^{-\Delta M \cdot x}.
\end{eqnarray}
Since $N_1$ is considered almost degenerate with $N_2$, we have $\Delta M \simeq 0$ and obtain
\begin{eqnarray}
\sigma_{\rm eff} |v_{\rm rel}|&=& \left(\frac{\sigma_{11}}{4} + \frac{\sigma_{12}}{2} + \frac{\sigma_{22}}{4}\right) |v_{\rm rel}| \nonumber \\
&=&  a_{\rm eff} + b_{\rm eff} v_{\rm rel}^2,
\end{eqnarray}
where
\begin{eqnarray}
a_{\rm eff}&=& \frac{a_{11}}{4}+\frac{a_{12}}{2}+\frac{a_{22}}{4},  \nonumber \\
b_{\rm eff}&=& \frac{b_{11}}{4}+\frac{b_{12}}{2}+\frac{b_{22}}{4}.
\label{Eq:aeff_beff}
\end{eqnarray}

The thermally averaged cross section can be written as $\langle \sigma_{\rm eff}|v_{\rm rel}| \rangle = a_{\rm eff} + 6b_{\rm eff}/x$ and the relic abundance of cold dark matter is estimated to be:
\begin{eqnarray}
\Omega h^2 = \frac{1.07\times 10^9 x_f}{g_\ast^{1/2} m_{\rm pl}({\rm GeV}) (a_{\rm eff}+3b_{\rm eff}/x_f )},
\end{eqnarray}
where $m_{\rm pl}=1.22\times 10^{19}$ ${\rm GeV}$, $g_{\ast} = 106.75$, and
\begin{eqnarray}
x_f = \ln \frac{0.038 g_{\rm eff} m_{\rm pl} M_1 \langle \sigma_{\rm eff} |v_{\rm rel}| \rangle}{g_\ast^{1/2} x_f^{1/2} }.
\end{eqnarray}
%

\section{Modified tribimaximal mixing \label{sec:tribi}}
In order to determine the magnitude of the elements of the Yukawa matrix
\begin{eqnarray}
Y=\left( 
\begin{array}{*{20}{c}}
Y_{e1} & Y_{e2} & Y_{e3} \\
Y_{\mu 1} & Y_{\mu 2} & Y_{\mu 3} \\
Y_{\tau 1} & Y_{\tau 2} & Y_{\tau 3} \\
\end{array}
\right),
\label{Eq:Y}
\end{eqnarray}
in Eq.(\ref{Eq:L_yukawa}), we employ methods ${\rm (c')}$ and ${\rm (d')}$ from the Introduction. 

According to method ${\rm (c')}$, assuming the mass matrix of the charged lepton is diagonal, we write the neutrino mixing matrix 
\begin{eqnarray}
U=\left( 
\begin{array}{*{20}{c}}
U_{e1} & U_{e2} & U_{e3} \\
U_{\mu 1} & U_{\mu 2} & U_{\mu 3} \\
U_{\tau 1} & U_{\tau 2} & U_{\tau 3} \\
\end{array}
\right)
\label{Eq:U}
\end{eqnarray}
as the following modified tribimaximal mixing with $\zeta=0$ \cite{Singirala2017CPC,Sruthilaya2015NJP}:
\begin{eqnarray}
U&=&\left( {\begin{array}{*{20}{c}}
\cos\theta & \sin\theta & 0\\
-\frac{\sin\theta}{\sqrt{2}} & \frac{\cos\theta}{\sqrt{2}} & \frac{1}{\sqrt{2}}\\
\frac{\sin\theta}{\sqrt{2}} & -\frac{\cos\theta}{\sqrt{2}} & \frac{1}{\sqrt{2}}\\
\end{array}} \right) \nonumber \\
&& \times \left( 
\begin{array}{*{20}{c}}
\cos\varphi&0&e^{-i\zeta}\sin\varphi \\
0&1 &0\\
-e^{i\zeta}\sin\varphi &0&\cos\varphi 
\end{array}
 \right).
 \label{Eq:UMTB}
\end{eqnarray}
The neutrino mixing angles $\theta_{12}$, $\theta_{23}$ and $\theta_{13}$ can be defined via the elements of the neutrino mixing matrix \cite{PDG}
\begin{eqnarray}
\sin^2\theta_{12}&=&\frac{|U_{e2}|^2}{1-|U_{e3}|^2}, \quad \sin^2\theta_{23}=\frac{|U_{\mu 3}|^2}{1-|U_{e3}|^2}, \nonumber \\
\sin^2\theta_{13}&=&|U_{e3}|^2.
\end{eqnarray}
We obtain 
\begin{eqnarray}
&& \sin^2\theta_{12}=0.336, \nonumber \\
&& \sin^2\theta_{23}=0.400, \nonumber \\
&& \sin^2\theta_{13}=0.0202, 
\label{Eq:mixing_angle_for_35_10}
\end{eqnarray}
for $\theta = 35^\circ$, $\varphi = 10^\circ$ which can accommodate the result of the following global fitting ($3 \sigma$) for the so-called normal mass ordering of neutrino masses \cite{Esteban2017JHEP}:
\begin{eqnarray}
&& \sin^2\theta_{12} = 0.271 - 0.345, \nonumber \\
&& \sin^2\theta_{23} = 0.385 - 0.635 , \nonumber \\
&& \sin^2\theta_{13} = 0.01934 - 0.02392.
\label{Eq:mixing_angle_3sigma}
\end{eqnarray}
Although the neutrino mass ordering (either the normal mass ordering or the inverted mass ordering) is not determined, a global analysis shows that the preference for the normal mass ordering is mostly due to neutrino oscillation measurements \cite{Salas2018arXiv,Salas2018PLB}. We assume the normal mass hierarchical spectrum for the neutrinos.

Using the relation
\begin{eqnarray}
U^T M_\nu U = {\rm diag}.(m_1,m_2,m_3)
\label{Eq:UTMU},
\end{eqnarray}
where $m_1$, $m_2$, and $m_3$ denote the neutrino mass eigenvalues, along with Eqs.(\ref{Eq:M_alpha_beta}) and (\ref{Eq:UMTB}), the vanishing off-diagonal elements of the mass matrix $M_\nu$ yield 
\begin{eqnarray}
Y=\left( 
\begin{array}{*{20}{c}}
Y_1 & Y_2 & Y_3 \\
-a_1 Y_1 & Y_2 & a_3 Y_3 \\
a_2 Y_1 & -Y_2 & a_4 Y_3 \\
\end{array}
\right),
\label{Eq:Y_mtb}
\end{eqnarray}
and the neutrino mass eigenvalues are obtained as
\begin{eqnarray}
m_1 = c_1 Y_1^2 \Lambda_1, \quad m_2 = c_2 Y_2^2 \Lambda_2, \quad m_3 = c_3 Y_3^2 \Lambda_3,
\label{Eq:m1m2m3}
\end{eqnarray}
where
\begin{eqnarray}
\Lambda_1 \simeq \Lambda_2,
\end{eqnarray}
and
\begin{eqnarray}
&& a_1= 0.647, \quad a_2=0.343, \quad  a_3=4.40, \quad a_4=5.39, \nonumber \\
&& c_1=1.54, \quad c_2=3.00, \quad c_3=49.4,
\label{Eq:values_a_1etc}
\end{eqnarray}
for $\theta = 35^\circ$ and $\varphi = 10^\circ$ \cite{Singirala2017CPC}. We note that the values in Eq.(\ref{Eq:values_a_1etc}) are different from those in Ref.\cite{Singirala2017CPC}. In Ref.\cite{Singirala2017CPC},  $\theta = 35^\circ$ and $\varphi = 12^\circ$ were taken from the neutrino oscillation pattern in Ref \cite{Forero2014PRD}. On the other hand, we take $\theta = 35^\circ$, $\varphi = 10^\circ$ from the global fitting data in Ref.\cite{Esteban2017JHEP}. This difference does not greatly affect our conclusions.

The squared mass differences of the neutrinos are given by
\begin{eqnarray}
\Delta m_{21}^2 &=& m_2^2-m_1^1= [(c_2 Y_2^2)^2 - (c_1 Y_1^2)^2] \Lambda_1^2, \nonumber \\
\Delta m_{31}^2 &=& m_3^2-m_1^1= (c_3 Y_3^2 \Lambda_3)^2 - (c_1 Y_1^2 \Lambda_1)^2, 
\label{Eq:a1a2a3a4c1c2c3}
\end{eqnarray}
and we obtain the relations
\begin{eqnarray}
Y_2^2 &=& \frac{1}{c_2\Lambda_1}\sqrt{\Delta m_{21}^2+ (c_1 Y_1^2\Lambda_1)^2}, \nonumber \\
Y_3^2 &=& \frac{1}{c_3\Lambda_3} \sqrt{\Delta m_{31}^2+ (c_1 Y_1^2\Lambda_1)^2}.
\label{Eq:Y2Y3}
\end{eqnarray}
The best-fit values of the squared mass differences are reported as \cite{Esteban2017JHEP}
\begin{eqnarray} 
&& \Delta m^2_{21} = 7.50 \times 10^{-5} {\rm eV}^2, \nonumber \\
&& \Delta m^2_{31} = 2.524 \times 10^{-3} {\rm eV}^2. 
\label{Eq:best_fit_deltams}
\end{eqnarray}

With the definition 
\begin{eqnarray} 
r_k=\frac{M_k}{m_0},
\label{Eq:r_k}
\end{eqnarray}
there are five parameters $\lambda, r_1, r_3, m_0, Y_1$ to calculate the relic abundance of dark matter and the branching ratio of the $\mu \rightarrow e \gamma$ process.

\section{One zero texture \label{sec:onezero}}
\subsection{Model parameters \label{subsec:onezero_param}}
According to method ${\rm (d')}$ in the Introduction, we assume that the flavor neutrino mass matrix $M_\nu$ has one zero element. There are six patterns for the flavor neutrino mass matrix $M_\nu$:
\begin{eqnarray}
&& {\rm G}_1:
\left( 
\begin{array}{*{20}{c}}
0 & \times & \times \\
- & \times & \times \\
- & - & \times \\
\end{array}
\right),
\quad
 {\rm G}_2:
\left( 
\begin{array}{*{20}{c}}
\times & 0 & \times \\
- & \times & \times \\
- & - & \times \\
\end{array}
\right),
\nonumber \\
&&{\rm G}_3:
\left( 
\begin{array}{*{20}{c}}
\times & \times &0 \\
- & \times & \times \\
- & - & \times \\
\end{array}
\right),
\quad
{\rm G}_4:
\left( 
\begin{array}{*{20}{c}}
\times & \times & \times \\
- & 0 & \times \\
- & - & \times \\
\end{array}
\right),
\nonumber \\
&& {\rm G}_5:
\left( 
\begin{array}{*{20}{c}}
\times & \times & \times \\
- & \times & 0 \\
- & - & \times \\
\end{array}
\right),
\quad 
{\rm G}_6:
\left( 
\begin{array}{*{20}{c}}
\times & \times & \times \\
- & \times & \times \\
- & - &0 \\
\end{array}
\right).
\label{Eq:G1G2G3G4G5G6}
\end{eqnarray}

For the ${\rm G_1}$ pattern, the relation
\begin{eqnarray}
M_{ee} =  Y_{e1}^2\Lambda_1 + Y_{e2}^2\Lambda_2 + Y_{e3}^2\Lambda_3 =0
\label{Eq:MeeG1}
\end{eqnarray}
is required by Eq.(\ref{Eq:M_alpha_beta}), where 
\begin{eqnarray}
\Lambda_k =  \frac{\lambda v^2}{16\pi^2}\frac{1}{m_0}\frac{r_k}{1-r^2_k}\left(1-\frac{r^2_k}{1-r^2_k}\ln\frac{1}{r^2_k} \right) .
\label{Eq:Lambda_k_r}
\end{eqnarray}
Since $\Lambda_k > 0$ for $r_k \neq 1$ and we assumed that $Y_{\alpha k}$ is real, Eq.(\ref{Eq:MeeG1}) yields $Y_{ek}=0$. However, the vanishing $Y_{ek}$ yields
\begin{eqnarray}
M_{e\mu} = \sum_{k=1}^3 Y_{ek}Y_{\mu k}\Lambda_k  =0, \nonumber \\
M_{e\tau} = \sum_{k=1}^3 Y_{ek}Y_{\tau k}\Lambda_k  =0,
\end{eqnarray}
as well as 
\begin{eqnarray}
\left( 
\begin{array}{*{20}{c}}
0 & 0 & 0 \\
- & \times & \times \\
- & - & \times \\
\end{array}
\right),
\end{eqnarray}
and the one-zero-texture assumption should be violated. Thus, the ${\rm G_1}$ pattern is excluded in the scotogenic model. Similarly, the ${\rm G_4}$ and ${\rm G_6}$ patterns are also excluded.

For the ${\rm G_2}$ pattern, the relation
\begin{eqnarray}
M_{e\mu} &=& Y_{e 1}Y_{\mu 1}\Lambda_1 + Y_{e 2}Y_{\mu 2}\Lambda_2 + Y_{e 3}Y_{\mu 3}\Lambda_3 \nonumber \\
&=& -a_1 Y_1^2 \Lambda_1 + Y_2^2 \Lambda_1 + a_3 Y_3^2 \Lambda_3 \nonumber \\
&=&0,
\end{eqnarray}
is required by Eqs.(\ref{Eq:M_alpha_beta}) and (\ref{Eq:Y_mtb}). Using Eq. (\ref{Eq:Y2Y3}), we have 
\begin{eqnarray}
&&-a_1 Y_1^2 \Lambda_1 + \frac{1}{c_2}\sqrt{\Delta m_{21}^2+ (c_1 Y_1^2\Lambda_1)^2} \nonumber \\
&& \quad + \frac{a_3 }{c_3} \sqrt{\Delta m_{31}^2+ (c_1 Y_1^2\Lambda_1)^2} =0,
\label{Eq:Y1forG2}
\end{eqnarray}
and $Y_1$ becomes a function of $\lambda, r_1, r_3$, and $m_0$. 

Similarly, we obtain 
\begin{eqnarray}
&&a_2 Y_1^2 \Lambda_1 - \frac{1}{c_2}\sqrt{\Delta m_{21}^2+ (c_1 Y_1^2\Lambda_1)^2} \nonumber \\
&& \quad + \frac{a_4}{c_3} \sqrt{\Delta m_{31}^2+ (c_1 Y_1^2\Lambda_1)^2} =0,
\label{Eq:Y1forG3}
\end{eqnarray}
for the ${\rm G_3}$ pattern and 
\begin{eqnarray}
&&-a_1 a_2 Y_1^2 \Lambda_1 - \frac{1}{c_2}\sqrt{\Delta m_{21}^2+ (c_1 Y_1^2\Lambda_1)^2} \nonumber \\
&& \quad  + \frac{a_3 a_4}{c_3} \sqrt{\Delta m_{31}^2+ (c_1 Y_1^2\Lambda_1)^2} =0,
\label{Eq:Y1forG5}
\end{eqnarray}
for the ${\rm G_5}$ pattern.

Thanks to the assumption of one zero texture for the flavor neutrino mass matrix, the number of parameters is reduced to four ($\lambda, r_1, r_3, m_0$) for the relic abundance of dark matter and the branching ratio of the $\mu \rightarrow e \gamma$ process. 

\subsection{Parameter dependence \label{subsec:onezero_param_dependence}}
We show the parameter dependence on the relic abundance of dark matter $\Omega h^2$ and the branching ratio ${\rm Br}(\mu \rightarrow e \gamma)$.

We can write Eq.(\ref{Eq:Lambda_k_r}) as
\begin{eqnarray}
\Lambda_k =  \frac{1}{m_0}\Lambda'_k(\lambda, r_k),
\label{Eq:Lambda_prime}
\end{eqnarray}
and obtain
\begin{eqnarray}
[(c_2 Y_2^2)^2 - (c_1 Y_1^2)^2] \Lambda^{'2}_1(\lambda, r_1) \propto m_0^2, \nonumber \\
(c_3 Y_3^2 )^2\Lambda^{'2}_3(\lambda, r_3) - (c_1 Y_1^2)^2 \Lambda^{'2}_1(\lambda, r_1)\propto m_0^2
\end{eqnarray}
from Eq.(\ref{Eq:a1a2a3a4c1c2c3}) and
\begin{eqnarray}
Y_1^2 \propto m_0,\quad Y_2^2 \propto m_0, \quad Y_3^2 \propto m_0
\label{Eq:YYsim_m_0}
\end{eqnarray}
from Eq.(\ref{Eq:Y2Y3}). Thus, $Y_{\alpha k}Y_{\beta k}$ is proportional to $m_0$:
\begin{eqnarray}
Y_{\alpha k}Y_{\beta k} =  m_0 Y'_{\alpha k}Y'_{\beta k},
\label{Eq:YYsim_m_0}
\end{eqnarray}
where $Y'_{\alpha k}$ is a function of $\lambda$ and $r_k$
\begin{eqnarray}
Y'_{\alpha k}=f(\lambda, r_k). 
\end{eqnarray}

The coefficients of the cross section in Eq.(\ref{Eq:a_b}) can be expressed in terms of $Y'_{\alpha k}$ as
\begin{eqnarray}
a_{ij}&=& \frac{1}{8\pi}\frac{r_1^2}{(r_1^2 +1)^2} \sum_{\alpha\beta}(Y'_{\alpha i} Y'_{\beta j} - Y'_{\alpha j} Y'_{\beta i})^2, \nonumber \\
b_{ij}&=&\frac{1-3r_1^2-r_1^4}{3(r_1^2+1)^2}a_{ij} \nonumber \\
&& +  \frac{1}{12\pi}\frac{r_1^2(r_1^4+1)}{(r_1^2+1)^4}  \sum_{\alpha\beta}Y'_{\alpha i} Y'_{\alpha j} Y'_{\beta i} Y'_{\beta j},
\end{eqnarray}
which are functions of $\lambda$ and $r_k$. Since the annihilation cross section is independent of $r_3$ [see Eq.(\ref{Eq:aeff_beff})], the relic abundance of dark matter depends on only $\lambda$ and $r_1$,
\begin{eqnarray}
\Omega h^2 = f(\lambda, r_1).
\label{Eq:omega_param}
\end{eqnarray}

On the other hand, the branching ratio ${\rm Br}(\mu \rightarrow e \gamma)$ depends on all four parameters $\lambda, r_1, r_3$ and $m_0$,
\begin{eqnarray}
{\rm Br}(\mu \rightarrow e \gamma)=f(\lambda, r_1, r_3, m_0).
\label{Eq:br_param}
\end{eqnarray}
%

\begin{figure}[t]
\begin{center}
\includegraphics{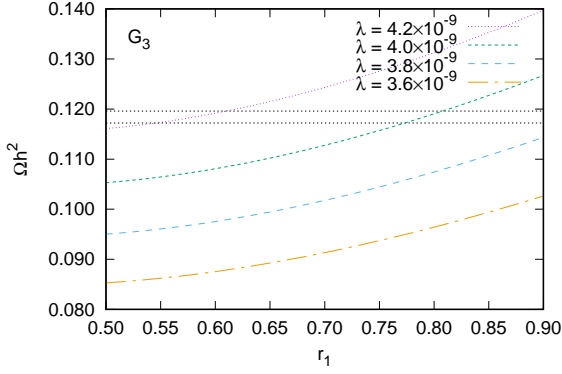}
\caption{The dependence of the relic abundance of dark matter $\Omega h^2$ on the mass ratio $r_1$ in the ${\rm G_3}$ pattern. The dotted horizontal lines show the upper and lower limits from observations.}
\label{fig:omega}
\end{center}
\end{figure}
\begin{figure}[t]
\begin{center}
\includegraphics{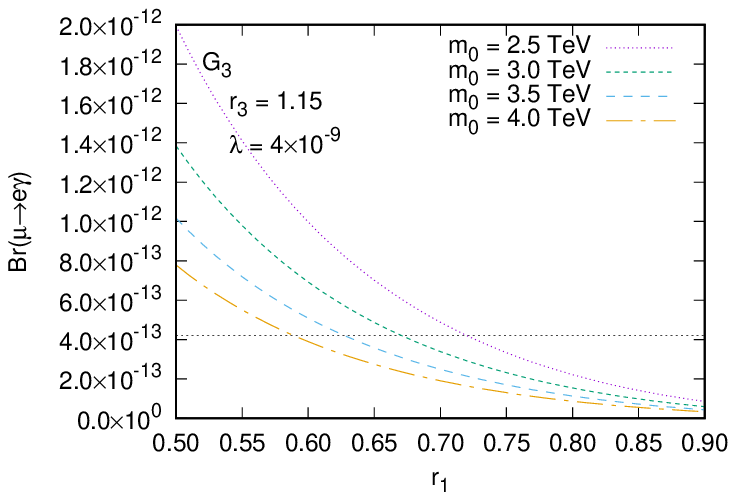}
\includegraphics{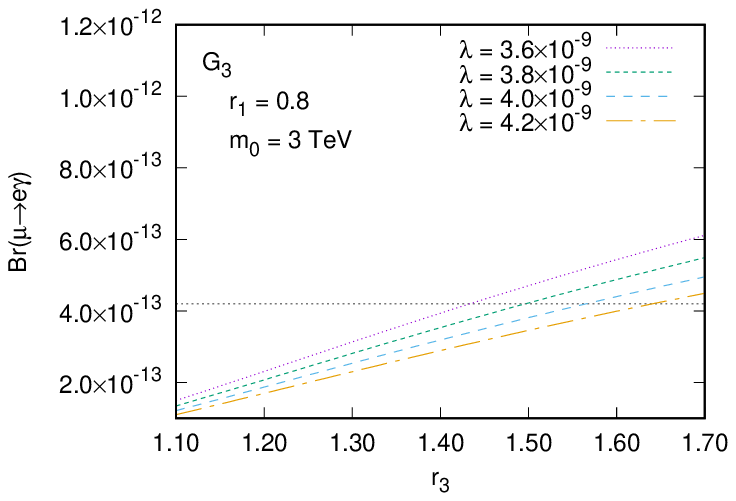}
\caption{The dependence of the branching ratio ${\rm Br}(\mu \rightarrow e \gamma)$ on the mass ratios $r_1$ (upper) and $r_3$ (lower). We take $r_3=1.15$ and $\lambda = 4 \times 10^{-9}$ in the upper panel while  $r_1=0.8$ and $m_0 = 3$ TeV in the lower panel. The dotted horizontal lines show the upper limits from observations.
}
\label{fig:br_r1_r3}
\end{center}
\end{figure}

\begin{figure}[t]
\begin{center}
\includegraphics{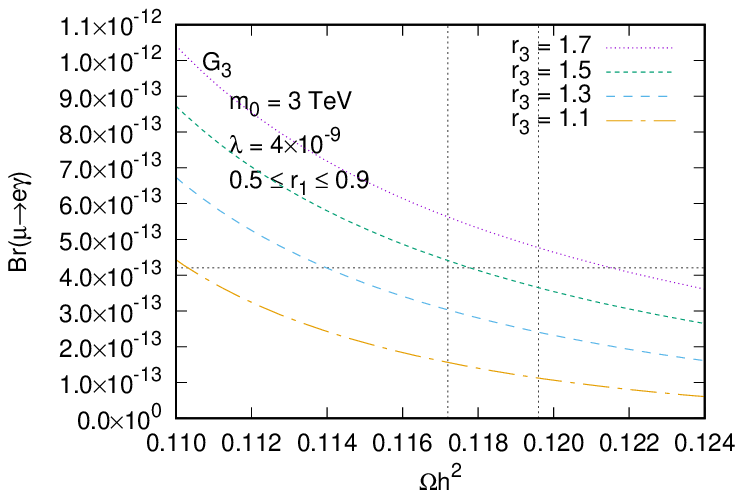}
\includegraphics{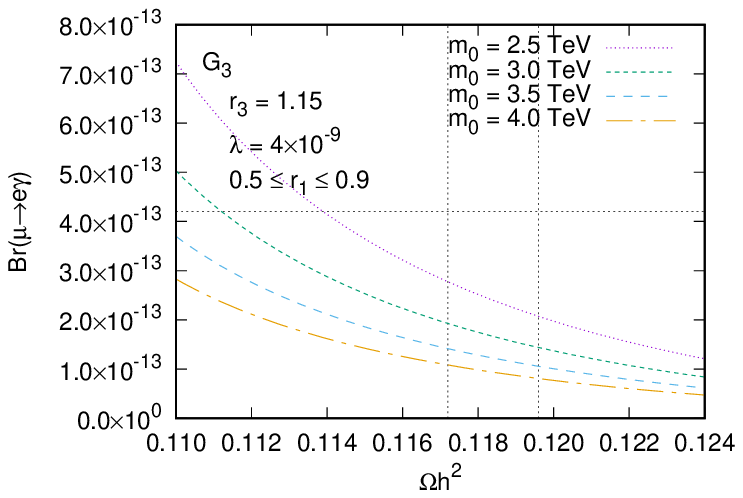}
\caption{The branching ratio ${\rm Br}(\mu \rightarrow e \gamma)$ vs the relic abundance of dark matter $\Omega h^2$ in the ${\rm G_3}$ pattern. In the upper panel $m_0=3$ TeV, $\lambda = 4 \times 10^{-9}$, and $0.5 \le r_1 \le 0.9$, while in the lower panel shows $r_3=1.15$, $\lambda = 4 \times 10^{-9}$, and $0.5 \le r_1 \le 0.9$. The dotted horizontal lines show the upper limits of ${\rm Br}(\mu \rightarrow e \gamma)$ and the dotted vertical lines show the lower and upper limits of $\Omega h^2$ from observations.}
\label{fig:br_omega}
\end{center}
\end{figure}

\subsection{${\rm G_3}$ \label{subsec:G3}}
We show that the ${\rm G_3}$ pattern within the scotogenic model is consistent with the observed neutrino oscillation data, the relic abundance of dark matter $\Omega h^2$, and the branching ratio ${\rm Br}(\mu \rightarrow e\gamma)$ from numerical calculations. 

First, to guaranty the consistency of the neutrino oscillation data, we take $\theta = 35^\circ$, $\varphi = 10^\circ$ and the best-fit values of the squared mass differences in Eq.(\ref{Eq:best_fit_deltams}). Next, we adopt the following standard criteria (see, for examples, Refs. \cite{one_loop_Kubo2006PLB,Ibarra2016PRD,Lindner2016PRD}). 1) The quartic coupling satisfies the relation $|\lambda| \ll 1$ for small neutrino masses. 2) Since we assumed that the additional lightest Majorana fermion $N_1$ is dark matter particle, we require $r_1 < r_3$. 3) The mass scale of new fields is a few TeV.  We take 
\begin{eqnarray}
3.6 \times 10^{-9} \le &\lambda& \le 4.2 \times 10^{-9}, \nonumber \\
0.5 \le &r_1& \le 0.99, \nonumber \\
1.1 \le &r_3& \le 3.0, \nonumber \\
2 {\rm TeV} \le &m_0& \le 4 {\rm TeV}.
\label{Eq:DMparameteter}
\end{eqnarray}

Let  us consider the benchmark parameter set
\begin{eqnarray}
\lambda=4 \times 10^{-9}, \ r_1 = 0.786, \ r_3 = 1.15, \ m_0=3 {\rm TeV}.
\label{Eq:benchmark}
\end{eqnarray}
Using these benchmark values, we obtain
\begin{eqnarray}
\Omega h^2 = 0.118, \quad {\rm Br}(\mu \rightarrow e\gamma) = 3.36 \times 10^{-13},
\end{eqnarray}
which are consistent with observations. The observed relic abundance is $\Omega h^2 = 0.1184 \pm 0.0012$ \cite{Planck2016AA}, while the measured upper limit of the branching ratio is $ {\rm Br}(\mu \rightarrow e\gamma) \le 4.2 \times 10^{-13}$ \cite{MEG2016arXiv}. Although the upper limits of the branching ratio of ${\rm Br}(\tau \rightarrow \mu \gamma) \le 4.4 \times 10^{-8}$ and ${\rm Br}(\tau \rightarrow e \gamma) \le 3.3 \times 10^{-8}$ were also reported \cite{BABAR2010PRL}, we only account for ${\rm Br}(\mu \rightarrow e\gamma)$ since it is the most stringent constraint. 

\begin{figure}[t]
\begin{center}
\includegraphics{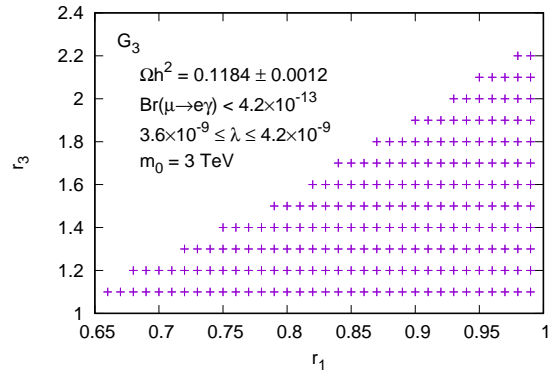}
\caption{Allowed region for $3.6 \times 10^{-9} \le \lambda \le 4.2 \times 10^{-9}$ and $m_0=3$ TeV in the $(r_1, r_3)$ plane satisfying the upper limit of the branching ratio of $\mu \rightarrow e \gamma$ and the dark matter relic abundance bounds in the ${\rm G_3}$ pattern. }
\label{fig:r1_r3}
\end{center}
\end{figure}

The results from a more general parameter search are shown in Figs.\ref{fig:omega}, \ref{fig:br_r1_r3}, \ref{fig:br_omega}, and \ref{fig:r1_r3}. 

Figure \ref{fig:omega} shows the dependence of the relic abundance of dark matter $\Omega h^2$ on the mass ratio $r_1$ in the ${\rm G_3}$ pattern. The dotted horizontal lines show the upper and lower limits from observations. The relic abundance of dark matter depends on only $\lambda$ and $r_1$ [see Eq.(\ref{Eq:omega_param})]. We see the existence of the allowed parameter set of $\{\lambda,r_1\}$ for the observed $\Omega h^2$.

Figure \ref{fig:br_r1_r3} shows the dependence of the branching ratio ${\rm Br}(\mu \rightarrow e \gamma)$ on the mass ratios $r_1$ (upper) and $r_3$ (lower). We take $r_3=1.15$ and $\lambda = 4 \times 10^{-9}$ in the upper panel, while $r_1=0.8$ and $m_0 = 3$ TeV in the lower panel. The dotted horizontal lines show the upper limits from observations. The ratio ${\rm Br}(\mu \rightarrow e \gamma)$ depends on all four parameters of $\lambda, r_1, r_3$, and $m_0$ [see Eq.(\ref{Eq:br_param})]. Figure \ref{fig:br_r1_r3} shows examples of allowed parameter set $\{ \lambda, r_1, r_3, m_0\}$.

Figure \ref{fig:br_omega} shows the branching ratio ${\rm Br}(\mu \rightarrow e \gamma)$ vs the relic abundance of dark matter $\Omega h^2$ in the ${\rm G_3}$ pattern. In the upper panel $m_0=3$ TeV, $\lambda = 4 \times 10^{-9}$, and $0.5 \le r_1 \le 0.9$, while in the lower panel shows $r_3=1.15$, $\lambda = 4 \times 10^{-9}$, and $0.5 \le r_1 \le 0.9$. The dotted horizontal lines show the upper limits of ${\rm Br}(\mu \rightarrow e \gamma)$ and the dotted vertical lines show the lower and upper limits of  $\Omega h^2$ from observations. We see the existence of the allowed parameter set $\{ \lambda, r_1, r_3, m_0\}$ for the observed $\Omega h^2$ and ${\rm Br}(\mu \rightarrow e \gamma)$.

Figure \ref{fig:r1_r3} depicts the allowed region for $3.6 \times 10^{-9} \le \lambda \le 4.2 \times 10^{-9}$ and $m_0=3$ TeV in the $(r_1, r_3)$ plane satisfying the upper limit of the branching ratio of $\mu \rightarrow e \gamma$ and the dark matter relic abundance bounds in the ${\rm G_3}$ pattern. 

We conclude that the ${\rm G_3}$ pattern within the scotogenic model is consistent with the observations.

\subsection{${\rm G_2}$ and ${\rm G_5}$ \label{subsec:G2G5}}
We show that the ${\rm G_2}$ and ${\rm G_5}$ patterns are not favorable for the scotogenic model with real Yukawa matrix elements. 

Because we assume that the Yukawa matrix elements are real, Eqs.(\ref{Eq:Y1forG3}) and (\ref{Eq:Y1forG5}) should have a real solution for $Y_1$. For the ${\rm G_2}$ pattern, the benchmark parameter set in Eq.(\ref{Eq:benchmark}) with $\theta = 35^\circ$ and $\varphi = 10^\circ$ yields the neutrino mixings shown in Eq.(\ref{Eq:mixing_angle_for_35_10}); however, it yields a complex Yukawa matrix element $Y_1=0.94 i$.  If we replace $\theta = 35^\circ$ with $\theta = 36^\circ$, we obtain a real Yukawa matrix element $Y_1=0.56$, but also obtain
\begin{eqnarray}
&& \sin^2\theta_{12} = 0.353, \nonumber \\
&& \sin^2\theta_{23} = 0.397, \nonumber \\
&& \sin^2\theta_{13} = 0.0197.
\end{eqnarray}
This value of $\sin^2\theta_{12}$ is out of the range of $3\sigma$ data in Eq.(\ref{Eq:mixing_angle_3sigma}). A similar result for the 
${\rm G_5}$ pattern is obtained.

We performed a scan of the parameter space for real $Y_1$ with the following sample points:
\begin{eqnarray}
\theta &=& \{34^\circ, 35^\circ, 36^\circ, 37^\circ \}, \nonumber \\
\varphi &=& \{9^\circ, 10^\circ, 11^\circ, 12^\circ \}, 
\end{eqnarray}
and
\begin{eqnarray}
\lambda &=& \{3.6, 3.8, 4.0, 4.2 \} \times 10^{-9}, \nonumber \\
r_1 &=& \{0.5,0.6,0.7,0.8,0.9 \}, \nonumber \\
r_3 &=& \{1.1,2.1,3.1 \}.
\end{eqnarray}
Since Eqs.(\ref{Eq:Lambda_prime}) and (\ref{Eq:YYsim_m_0}) are satisfied, $Y_1^2 \Lambda_1$ is independent of $m_0$. Equations (\ref{Eq:Y1forG3}) and (\ref{Eq:Y1forG5}) are also independent of $m_0$.

We obtain
\begin{eqnarray}
&& \sin^2\theta_{12} = 0.352 - 0.372, \nonumber \\
&& \sin^2\theta_{23} = 0.374 - 0.406, \nonumber \\
&& \sin^2\theta_{13} = 0.0156 - 0.0283
\end{eqnarray}
for the ${\rm G_2}$ pattern and 
\begin{eqnarray}
&& \sin^2\theta_{12} = 0.352 - 0.372, \nonumber \\
&& \sin^2\theta_{23} = 0.374 - 0.397, \nonumber \\
&& \sin^2\theta_{13} = 0.0192 - 0.0283
\end{eqnarray}
for the ${\rm G_5}$ pattern from the scan. These predicted values of $\sin^2\theta_{12}$ are out of the range of $3\sigma$ data. 

We conclude that the ${\rm G_2}$ and ${\rm G_5}$ patterns are not favorable for the scotogenic model with real Yukawa matrix elements. 

\section{\label{sec:summary}Summary}
We have assumed that the neutrino mixing is described by a modified tribimaximal mixing and the Yukawa matrix elements are real. Moreover, we have required the flavor neutrino mass matrix to have one zero element. There are six patterns of the flavor neutrino mass matrix, ${\rm G_1}$, ${\rm G_2}$, $\cdots$, ${\rm G_6}$ in the one-zero scheme. 

It turned out that only one pattern, ${\rm G_3}$,  within the scotogenic model is consistent with the observed neutrino oscillation data, the relic abundance of dark matter, and the upper limit of the branching ratio of the $\mu \rightarrow e \gamma$ process. For three patterns (${\rm G_1}, {\rm G_4}$ and ${\rm G_6}$), the texture zero assumption should be violated. Two patterns (${\rm G_2}$ and ${\rm G_5}$) are not favorable because the predicted $\sin^2\theta_{12}$ is out of the range of $3\sigma$ data. 

Finally, we would like to comment on whether the result in this paper is robust in the presence of $CP$ violation. Since we assumed that all elements of the Yukawa matrix are real, there is no $CP$-violating source in the Yukawa sector. If we had included $CP$-violating phases such as in the realistic tribimaximal neutrino mixing patterns in Ref \cite{Chen2018PRD}, the results may have been different. A detailed analysis of this topic will be found in our future study.  

\vspace{3mm}






\end{document}